\documentclass{Interspeech2024}
\usepackage{xcolor}

\interspeechcameraready

\let\OLDthebibliography\thebibliography
\renewcommand\thebibliography[1]{
  \OLDthebibliography{#1}
  \setlength{\parskip}{1.2pt}
  \setlength{\itemsep}{1.5pt plus 0.5ex}
}

\title{Language-Universal Speech Attributes Modeling for Zero-Shot Multilingual Spoken Keyword Recognition}
\name[affiliation={1}]{Hao}{Yen}
\name[affiliation={1}]{Pin-Jui}{Ku}
\name[affiliation={1,2,3}]{Sabato Marco}{Siniscalchi}
\name[affiliation={1}]{Chin-Hui}{Lee}

\address{
  $^1$Georgia Institute of Technology, U.S.A\\\
  $^2$Università degli Studi di Palermo, Italy \\
  $^3$Norwegian University of Science and Technology, Norway}
\email{\{rick.yen, pku9\}@gatech.edu}


\begin{document}

\maketitle

\begin{abstract}
\vspace{-.1cm}
We propose a novel language-universal approach to end-to-end automatic spoken keyword recognition (SKR) leveraging upon (i) a self-supervised pre-trained model, and (ii) a set of universal speech attributes (manner and place of articulation). Specifically, Wav2Vec2.0 is used to generate robust speech representations, followed by a linear output layer to produce attribute sequences. A non-trainable pronunciation model then maps sequences of attributes into spoken keywords in a multilingual setting. Experiments on the Multilingual Spoken Words Corpus show comparable performances to character- and phoneme-based SKR in seen languages. The inclusion of domain adversarial training (DAT) improves the proposed framework, outperforming both character- and phoneme-based SKR approaches with 13.73\% and 17.22\% relative word error rate (WER) reduction in seen languages, and achieves 32.14\% and 19.92\% WER reduction for unseen languages in zero-shot settings.
\end{abstract}

\vspace{-.1cm}
\section{Introduction}
\label{sec:intro}
Spoken Keyword Recognition (SKR) is an important sub-task within automatic speech recognition (ASR). SKR aims to identify a specific target keyword out of a set of predefined candidates~\cite {Zeppenfeld1992, Rohlicek1989}, also known as speech command recognition. Owing to its wide applicability to various domains, such as smart home devices~\cite{Bahpai2019} or crime defections~\cite{Kavya2014}, SKR has long been an important research topic in the speech processing field~\cite{Dinushika2019, Seo2021}. Moreover, the rising popularity of AI-assisted devices has sparked a heightened interest in SKR for customizable user experiences. However, a common requirement to build a high-performance SKR system is to prepare extensive labeled training data, which is often impractical in real-world scenarios. In fact, it is generally favorable to build a language-universal SKR system that can handle to any input or language, particularly in zero-shot transfer~\cite{Xian2017} or low-resource scenarios~\cite{Besacier2014}.

Recent studies have treated SKR as a classification task~\cite{Andrade2018, majumdar2020matchboxnet, berg21_interspeech}, which directly associates audio speech with specific labels. These ASR-free systems have a fixed output dimension that corresponds to the number of the predefined keywords and cannot be altered during testing, limiting their ability to handle out-of-domain scenarios like out-of-vocabulary words or unseen languages. Another category of approach adopts an ASR-based systems~\cite{Dinushika2019, Shahamiri2014}, but these typically rely on high-level tokens, such as characters and phonemes. The inherent disadvantage of using those tokens is that they are language-dependent, which means those systems are unable to process tokens from different languages, or keywords that have not been encountered during training. Hence, it is crucial for researchers to explore knowledge sharing among multiple languages, for example, defining a universal set of acoustic-phonetic units that work for multiple or even for all languages, in order to build a language-universal SKR system.

Speech attributes are unique features that explain the production of speech sounds by the mouth's articulators and are consistent across all languages. Opting for an attribute-based language characterization provides two main benefits over traditional high-level tokens. Firstly, these attributes are universally defined, removing the need to increase the attribute token set or alter models when introducing new languages. Additionally, these attributes can be updated and improved as new linguistic data is collected, regardless of the languages. Previous works on attribute modeling focused on automatic speech attribute transcription (ASAT)~\cite{Lee2013}, which built and utilized various attribute detectors with success for several speech applications~\cite{Siniscalchi2012,Chen2014,Li2016Detecting}. Recently, phonetic knowledge and speech attributes have been introduced to the task of multilingual ASR. However, most of these works focused on the recognition of characters, phonemes, or syllables~\cite{Yen2024,Lee2023,Glocker2023,li2019zeroshot,Li2019}, which should be considered as an initial step to build a multilingual ASR as the final word recognition is of most importance in the speech recognition task. Moreover, some of the previous works~\cite{Yen2024,Lee2023,Glocker2023} still rely on character and phoneme information to make the final predictions. These approaches typically view speech attributes as complimentary constraints instead of using them explicitly for the final recognition.

In this work, we extend previous universal attribute modeling efforts, from phoneme to word recognition, and present a system toward building a multilingual E2E ASR. The design concept of the proposed system consists of a comprehensive inventory of fundamental attribute units, a pre-trained Wav2Vec2.0 encoder, a linear output layer, and a non-trainable pronunciation model. The encoder initially generates robust features, which are then fed as input to a linear output layer, which models articulatory characteristics directly by generating sequences of posteriors for each attribute token. A non-trainable pronunciation model is adopted to map such sequences to the most likely keywords. To further enhance the language-universality, a domain adversarial training (DAT)~\cite{Ganin2017} technique is integrated into our proposed system, aiming to assist the pre-trained encoder in learning language-invariant features. Our experiments demonstrate that attribute-based system are more universal across languages than character- and phoneme-based systems. In addition, through modeling universal speech attributes, our proposed system successfully achieves zero-shot recognition for out-of-vocabulary keywords, referring to keywords not included in the predefined training vocabulary, and unseen languages, i.e., languages not represented during training. 


\vspace{-.1cm}
\section{Related Work}
\label{sec:related}
\vspace{-.1cm}

\subsection{Language-Universal Multilingual Modeling}
\vspace{-.05cm}
Multilingual ASR involves simply taking a union of the token sets (e.g., phoneme, character, subword, or even byte) of all languages to form the output token set and subsequently training an E2E model using all available data~\cite{Lin2009, Watanabe2017, Toshniwal2017}. Such an E2E model enables it to recognize speech in any language that was included in its training data. However, sharing the output token space can inevitably present a new challenge for high-level tokens such as characters and phonemes. While certain characters might be common across multiple languages, it is important to recognize that the same character can have varying pronunciations depending on the language. This variability in pronunciation implies that characters are not suitable for universal multilingual modeling. The use of phonemes, despite the fact that they can be universally defined with International Phonetic Association (IPA)~\cite{Smith2000}, faces the challenges as each language may assign different symbols for the same IPA token. This inconsistency in the definition of phonemes among languages suggests that they are also not an ideal choice as universal token.

Recently, several studies have been conducted to tackle multilingual modeling using universal speech units. In~\cite{Li2020,Yan2021}, the authors incorporate a universal narrow phone set, called \textit{allophone} to build phone-based recognition models. Though \textit{allophone} can be considered as fundamental units, the total size of units is not compact and thus posing challenges in modeling when large amount of languages present. Moreover, the performance is dependent on the clarity of allophone-phoneme dynamics in the selected training languages.

\vspace{-.05cm}
\subsection{Wav2Vec2.0}
\label{sec:w2v2}
\vspace{-.05cm}
Wav2vec2.0 is a self-supervised learning architecture that learns meaningful latent representation from raw speech data. In this framework, the speech input is first processed through a series of convolutional neural network (CNN) layers, each layer designed to capture different aspects of audio signal. Upon the extraction of these latent features, a Transformer encoder network refines these representations by embedding a richer contextual information into the data. The training for Wav2vec2.0 is similar to the strategies employed by masked language models such as BERT~\cite{Devlin2019}. It involves the model learning to accurately predict the original latent representations using the context provided by the other contextualized representations. This prediction task forces the model to develop an understanding of the speech that is deeply rooted in the context provided by the surrounding audio. Wav2vec2.0 models can be trained on large multilingual corpora and can then be used as a pre-trained model for finetuning on other target dataset.

\vspace{-.05cm}
\subsection{Domain Adversarial Training (DAT)}
\label{sec:dat}
\vspace{-.05cm}
Domain adversarial training is a powerful technique for improving the adaptability of machine learning models, especially in minimizing discrepancies between the training and testing data. The idea is to learn features to that are consistent across different domains with the help of adversarial training~\cite{Goodfellow2014}. Such domain-invariant features ensure that models can retain high performance even when deployed in environments different from their original training conditions. Recently, DAT is extensively used in enhancing the robustness of ASR systems across varied conditions, including speakers~\cite{Meng2018,Meng2019,Luu2020}, accents~\cite{Sun2018,Hu2021}, noises~\cite{Beutel2017}, and languages~\cite{Hu2019AdversarialTF}.

\begin{figure}[t!]
    \centering
    \includegraphics[width=0.8\columnwidth]{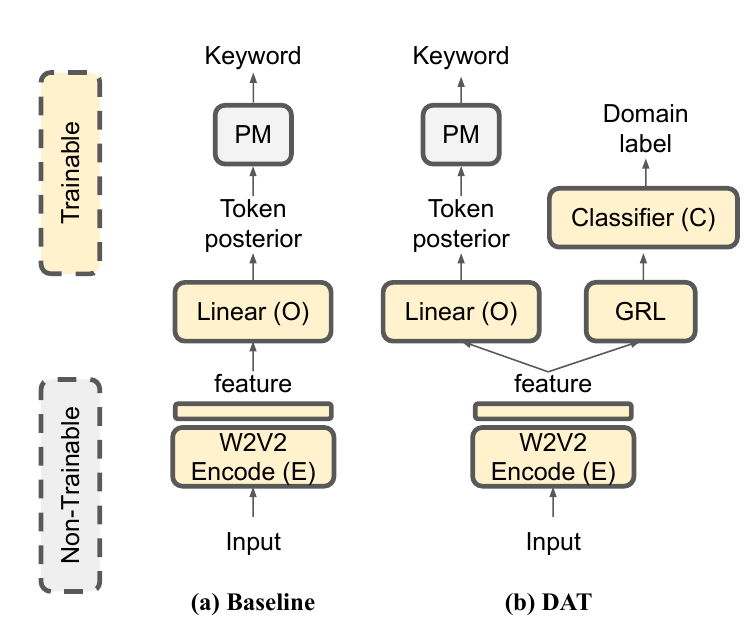}
    \caption{Frameworks studied in this work. "W2V2" refers to the Wav2Vec2.0 pre-trained model, and "PM" refers to pronunciation model. In (a), the Baseline system is trained from scratch with different tokens, namely characters, phonemes and speech attributes. In (b), a domain adversarial training (DAT) framework. GRL refers to gradient reversal layer and Classifier denotes a language classifier.}
    \label{fig:framework}
    \vspace{-.4cm}
\end{figure}

\section{Proposed Language-Universal Spoken Keyword Recognition}
\label{sec:propose}

\vspace{-.05cm}
\subsection{ASR-based Spoken Keyword Recognition System}
\label{sec:ASR}
\vspace{-.05cm}
Figure~\ref{fig:framework}(a) illustrates our proposed ASR-based SKR system, which comprises a pre-trained Wav2Vec2.0 encoder, a Linear output layer, and a non-trainable pronunciation model (PM). The encoder first produces the input features and feeds them to the Linear output layer, which aims to predict sequences of token posteriors. The system is trained under full-supervision with corresponding label tokens via the CTC criterion. The non-trainable pronunciation model consists of a lexicon with the collection of all the keywords and their corresponding token sequences in characters, phonemes, or attributes. Given the predicted token posteriors, PM applies a simple beam search to obtain the most likely keyword. The ASR-based system architecture has the capability to generalize to out-of-domain scenarios, such as out-of-vocabulary keywords and unseen languages, due to its ability to generate arbitrary sequences of tokens, and then rely on the provided lexicon to perform recognition.

\vspace{-.05cm}
\subsection{Universal Speech Attribute Modeling}
\label{sec:msam}
\vspace{-.05cm}
The set of speech attributes used in this work are acoustic-phonetic features, namely, seven \textit{manner of articulation} (M) classes (approximant, tap, fricative, affricate, nasal, stop, and vowel), and ten \textit{place of articulation} (P$_\text{c}$) (bilabial, labiodental, dental, alveolar, postalveolar, retroflex, palatal, velar, uvular, glottal). As pointed out in~\cite{Tang2003}, a difficulty in using manner and place of articulation for ASR applications is that vowels and consonants cannot be mapped into a common linguistic space, because place of articulation has been differently defined for them. Therefore, in this study, we define a different category $P_v$, which consists of eight classes (high, semi-high, upper-mid, mid, lower-mid, semi-mid, low, and unknown) to divide the vowels based on how they are pronounced. The inventory of universal attribute units adopted in this paper is shown in Table~\ref{tab:attributes}. Phonemes are categorized into each attribute based on the International Phonetic Association (IPA)~\cite{Smith2000} and PHOIBLE~\cite{phoible}, a large database of phone inventories for more than 2000 languages. Each attribute token is a combination of manner and place of articulation. For instance, the consonant /r/ consists of articulatory attributes tap and alveolar, so will be assigned the token "tap-alveolar"; whereas the vowel /i/ will be assigned "vowel-high". By doing so, we significantly makes the number of tokens more compact, resulting in more training data for each token. In our experiments, there are a total of 114 character tokens, 73 phoneme tokens, and only 33 attribute tokens. Therefore, the attribute-based solution is expected to capture most of the distinct features that describe human sounds. The proposed universal approach can naturally support zero-shot transfer. This is possible because all phonemes can be categorized in one of the attributes based on Table~\ref{tab:attributes}. Even if we encountered unseen phonemes from different languages, we can still convert it into attribute tokens and expect the model to infer from those that share the similar acoustic-phonetic features.

\begin{table}[t!]
\vspace{-.5cm}
    \centering
    \caption{Inventory of universal speech attributes.}
    \label{tab:attributes}
    \begin{adjustbox}{width=0.9\columnwidth}
    \begin{tabular}{c|l}
        \toprule
        \textbf{Category} & \textbf{Attributes} \\ \midrule \midrule
         Manner (M) &  approximant, tap, fricative, affricate, nasal, stop, vowel \\ \midrule
         Place (P$_\text{c}$) & \makecell[tl]{bilabial, labiodental, dental, alveolar, postalveolar, \\  retroflex, palatal, velar, uvular, glottal} \\ \midrule
         Place (P$_\text{v}$) & \makecell[tl]{high, semi-high, upper-mid, mid, lower-mid, semi-mid, \\ low, unknown} \\ \bottomrule
    \end{tabular}
    \end{adjustbox}
    \vspace{-.4cm}
\end{table}

\vspace{-.05cm}
\subsection{DAT for Universal Spoken Keyword Recognition}
\vspace{-.05cm}
As depicted in Figure~\ref{fig:framework}(b),  The DAT framework we propose consists of a language-invariant encoder $E$, linear output layer $O$, and a language classifier $C$. During training, a gradient reversal layer (GRL) is used to minimize the ability of the encoder to distinguish languages. In other words, the output features from the encoder $E$ are expected to be language-invariant representations. The losses for $O$ and $C$ are denoted as $\mathcal{L}_O$ and $\mathcal{L}_C$, and the respective parameters of $E$, $O$, $C$ as $\theta_E$, $\theta_O$ , $\theta_C$. We update each weight following specified gradient descent rules,
\begin{equation}
\begin{aligned}
\theta_E & \leftarrow \theta_E-\alpha\left(\frac{\partial \mathcal{L}_O}{\partial \theta_E}-\lambda \frac{\partial \mathcal{L}_C}{\partial \theta_E}\right) \\
\theta_C & \leftarrow \theta_C-\alpha \frac{\partial \mathcal{L}_C}{\theta_C} \\
\theta_O & \leftarrow \theta_O-\alpha \frac{\partial \mathcal{L}_O}{\theta_O}
\end{aligned}
\end{equation} where $\alpha$ is the learning rate and $\lambda$ is the scale of $L_C$ gradients.

Integrating DAT into our system presents a clear benefit. Given that collecting exhaustive language data is impractical, models often rely on language information and cues during recognition. This reliance could compromise their capability to handle previously unseen keywords and languages. DAT helps reduce this dependency by minimizing language information in the encoder's features, promoting better generalization to new scenarios. Since our proposed framework models speech attributes, which are expected to be relatively invariant to languages in nature, DAT can be complimentary to eliminate the influence of any residual language information in the features.

\begin{table}[t]
\vspace{-.5cm}
    \centering
    \caption{Training and testing data size in number of words and samples for each language, including in-domain in-vocabulary (\textbf{ID-IV}), in-domain out-of-vocabulary (\textbf{ID-OOV}), and unseen languages (\textbf{UL}).}
    \vspace{-.2cm}
    \label{tab:dataset}
    \begin{adjustbox}{width=0.9\columnwidth}
    \begin{tabular}{lllrrr}
        \midrule
        \textbf{Set} & \textbf{Language} & \textbf{Family} & \textbf{\# words} & \textbf{\# train} & \textbf{\# test} \\
        \midrule
        \multirow{8}{*}{\textbf{ID-IV}} & English (en) & Germanic & 1755 & 1194255 & 150238 \\
        & German  (de) & Germanic & 602  & 403101  & 50799  \\
        & French  (fr) & Romance  & 376  & 253913  & 31828  \\
        & Persian (fa) & Iranian  & 330  & 211875  & 26637  \\
        & Spanish (es) & Romance  & 229  & 149488  & 18729  \\
        & Russian (ru) & Slavic   & 97   & 62324   & 7857   \\
        & Italian (it) & Romance  & 78   & 50569   & 6358   \\
        & Polish  (pl) & Slavic   & 62   & 39513   & 5047   \\ \midrule \midrule
        \multirow{3}{*}{\textbf{ID-OOV}} & Russian (ru) & Slavic   & 30   & -   & 1698   \\
        & Italian (it) & Romance  & 30   & -  & 4481   \\        
        & Polish  (pl) & Slavic   & 30   & -  & 2448    \\ \midrule
        \multirow{3}{*}{\textbf{UL}} & Turkish (tr) & Turkic   & 30   & 12801   & 1620   \\
        & Latvian (lv) & Baltic   & 30   & 2845   & 389   \\        
        & Lithuanian (lt) & Baltic & 30  & 4137   & 330    \\
        \bottomrule
        \end{tabular}
        \end{adjustbox}
        \vspace{-.4cm}
\end{table}

\vspace{-.1cm}
\section{Experiments and Results}
\label{sec:exp}
\vspace{-.1cm}

\vspace{-.05cm}
\subsection{Datasets}
\label{sec:dataset}
\vspace{-.05cm}
As shown in Table~\ref{tab:dataset}, we take into account eight rich-resource languages from Multilingual Spoken Words Corpus (MSWC)~\cite{Mazumder2021} as our in-domain in-vocabulary (\textbf{ID-IV}) set, namely English, German, French, Persian, Spanish, Russian, Italian, and Polish. \textbf{ID-IV} set refers to languages and keywords that have been seen and included during training. We select languages from multiple language families (a group of languages that are related through a common ancestral language), and keywords that have more than 500 training samples for each language. For zero-shot transfer, we consider two scenarios: (i) an in-domain out-of-vocabulary (\textbf{ID-OOV}) set, (ii) and an unseen language (\textbf{UL}) set. \textbf{ID-OOV} set denotes keywords that belong to one of the eight seen languages but not included during training. We choose the three in-domain languages that have the fewest training data, namely Russian, Italian, and Polish, and select 30 most frequent keywords that are not included in the \textbf{ID-IV} set as target keywords. The \textbf{UL} set, referring to languages not represented in the training set, includes three low-resource languages, Turkish, Latvian, and Lithuanian, which all belong to different language families than the eight \textbf{ID-IV} languages, since we want to consider more extreme scenarios. 

While character sequences can be obtained from each keyword itself, non-aligned phoneme sequences are obtained by running the open-source tool phonemizer\footnote{\url{https://github.com/bootphon/phonemizer}}. After processing all keywords and removing some of those containing ambiguous phoneme-based tokens, the resulting training and testing datasets, and the corresponding number of keywords and samples for each language are shown in Table~\ref{tab:dataset}. After obtaining the phoneme sequences, we then use the mapping relationship described in Section~\ref{sec:msam} to transform each keyword into a sequence of attributes.

\vspace{-.05cm}
\subsection{Experimental Settings}
\vspace{-.05cm}
We use a publicly available pre-trained self-supervised model as our encoder model for all experiments, namely Wav2Vec2.0~\cite{baevski2020wav2vec}. We choose the "base" architecture - 95M parameters leveraging  12 Transformer encoder layers, with embedding size of 768 and 8 attention heads.  It is pre-trained on 960 hours of unlabeled audio from LibriSpeech dataset~\cite{Vassil2015}. The language classifier in DAT framework consists of 3 linear layers with the hidden dimension of 1024 and predicts eight in-domain language classes as described in Section~\ref{sec:dataset}. AdamW optimizer~\cite{Loshchilov2018} with $\beta_1=0.9$ and $\beta_2=0.98$ is used in training all models, and early stopping criterion is used to terminate the training phase.


\vspace{-.05cm}
\subsection{DAT Results: Language Identification}
\vspace{-.05cm}
Table~\ref{tab:lid} presents the language identification accuracy for systems trained with and without Domain Adversarial Training (DAT). In the absence of DAT, our models retain the same architecture depicted in Figure\ref{fig:framework}(b), except for the exclusion of the gradient reversal layer (GRL). The removal of the GRL implies that the feature extractor does not aim to mislead the language classifier. Thus the feature extractor learns language-specific features to optimize recognition performance.

The results clearly show that  DAT leads to a significant decrease in language identification accuracy, confirming the encoders' effectiveness in minimizing language-specific information, and further strengthen our motivation in its use in an out-of-domain scenarios. Furthermore, our attribute modeling approach, when combined with DAT, further lowers language identification accuracy. This suggests a stronger capability for feature universality, as characters and phonemes are inherently more language-dependent, making the extraction of language-invariant features more challenging. 

\begin{table}[t!]
\vspace{-.3cm}
    \centering
    \caption{Language identification accuracy (\%) with (w/) and without (w/o) DAT for characters, phonemes and attributes.}
    \vspace{-.2cm}
    \label{tab:lid}
    \begin{adjustbox}{width=\linewidth}
    \begin{tabular}{c|ccc}
    \toprule
     System ($\downarrow$)/Units ($\rightarrow$)& Characters & Phonemes & Attributes (ours)\\
     \hline
    w/o DAT & 91.35 & 91.10 & 90.47 \\
    w/ DAT & 45.24 & 49.66 & {\bf 34.10} \\  \bottomrule
    \end{tabular}
    \end{adjustbox}
\end{table}

\begin{table}[t!]
\centering
\caption{Testing WER (\%) of the in-domain set on 8 rich-resource languages and the average (Avg.).}
\label{tab:id_results}
\begin{adjustbox}{width=\columnwidth}
\begin{tabular}{c|ccccccccc}
\toprule
\multirow{2}{*}{\textbf{System}} & \multicolumn{9}{c}{\textbf{ID-IV}} \\
   & en         & de         & fr             & fa         & es             & ru         & it             & pl             & Avg.               \\ \midrule
$\text{Base}_{char}$   & 13.14      & 12.57      & 13.65   & 14.77     & 11.89  & 15.50      & \textbf{14.39} & 15.00 & 13.86            \\ 
$\text{Base}_{phone}$   & \textbf{12.73} & \textbf{11.70} & \textbf{13.29}  & 14.30 & \textbf{11.84}      & \textbf{12.96} & 14.63      & 15.14    & \textbf{13.32}     \\ 
$\text{Base}_{attr}$ (ours)   & 13.28          & 12.04       & 13.69      & \textbf{13.53}          & 12.20          & 13.67      & 15.13         & \textbf{14.98}        & 13.56     \\ \midrule \midrule
$\text{DAT}_{char}$   & 18.73      & 16.45          & 19.55      & 16.91      & 14.99          & 17.98          & 17.19     & 18.62         & 17.55            \\ 
$\text{DAT}_{phone}$   & 17.76      & 18.33         & 21.12       & 20.00      & 17.58          & 14.97          & 18.91      & 17.65         & 18.29            \\
$\text{DAT}_{attr}$ (ours)   & \textbf{15.51} & \textbf{13.74} & \textbf{16.27} & \textbf{15.47} & \textbf{13.85} & \textbf{13.83} & \textbf{16.73} & \textbf{15.75} & \textbf{15.14}  \\ \bottomrule
\end{tabular}
\end{adjustbox}
\vspace{-.2cm}
\end{table}

\vspace{-.05cm}
\subsection{Recognition Results}
\vspace{-.05cm}
Three different sets of units, namely characters, phonemes and attributes, are modeled and compared using the architectures in Figure~\ref{fig:framework}. Although accuracy is commonly used in  speech command classification, we report Word Error Rate (WER) to align with standard measurement in ASR for each system. The results of the Baseline training, denoted as $\text{Base}_{char}$, $\text{Base}_{phone}$, and $\text{Base}_{attr}$, stands for training the systems for scratch on the training set. The results of DAT-based systems are denoted as $\text{DAT}_{char}$, $\text{DAT}_{phone}$, and $\text{DAT}_{attr}$ respectively.

\vspace{-.05cm}
\subsubsection{In-domain Results: Seen Languages}
\vspace{-.05cm}
Table~\ref{tab:id_results} summarizes the WER for the eight in-domain in-vocabulary (\textbf{ID-IV}) languages for both Baseline and DAT settings. For Baseline, all systems, namely $\text{Base}_{char}$, $\text{Base}_{phone}$, and $\text{Base}_{attr}$  have similar results. Since characters and phonemes are relative language-dependent, we argue that during Baseline training, those system rely more on the language information to perform recognition. In fact, after removing language information via DAT and making the features language-invariant, our proposed framework, $\text{DAT}_{attr}$, consistently outperforms $\text{DAT}_{char}$ and $\text{DAT}_{phone}$ for eight in-domain languages, with a relative word error rate (WER) reduction of 13.73\%, and 17.22\%, respectively. The results provide strong support for our argument that speech attributes are significantly more universal across languages than traditional tokens.

\begin{table}[t!]
\vspace{-.3cm}
\centering
\caption{Zero-shot transfer: Testing WER (\%) of the 3 in-domain out-of-vocabulary (\textbf{ID OOV}) keywords from Russian, Italian, and Polish.}
\vspace{-.2cm}
\label{tab:idoov_results}
\begin{adjustbox}{width=0.7\columnwidth}
\begin{tabular}{c|cccc}
\toprule
\multirow{2}{*}{\textbf{System}}  & \multicolumn{4}{c}{\textbf{ID OOV}} \\
  & ru         & it             & pl             & Avg.          \\ \midrule
$\text{Base}_{char}$            & 63.96      & 44.50      & 40.77    & 49.74          \\ 
$\text{Base}_{phone}$           & \textbf{31.15}      & \textbf{40.62}      & 33.29    & \textbf{35.02}         \\ 
$\text{Base}_{attr}$ (ours)     & 31.57      & 41.89      & \textbf{32.80}    & 35.42      \\ \midrule \midrule
$\text{DAT}_{char}$             & 54.89      & 40.19      & 37.95         & 44.34          \\ 
$\text{DAT}_{phone}$            & \textbf{28.15}      & \textbf{38.61}      & 34.11         & \textbf{33.62}          \\ 
$\text{DAT}_{attr}$ (ours)      & 29.98      & 40.28      & \textbf{30.23} & 33.81     \\ \bottomrule 
\end{tabular}
\end{adjustbox}
\end{table}

\begin{table}[t!]
\centering
\caption{Zero-shot transfer: Testing WER (\%) of the 3 unseen languages (\textbf{UL}), namely Turkish, Latvian, and Lithuanian.}
\vspace{-.2cm}
\label{tab:ul_results}
\begin{adjustbox}{width=0.7\columnwidth}
\begin{tabular}{c|cccc}
\toprule
\multirow{2}{*}{\textbf{System}}  & \multicolumn{4}{c}{\textbf{UL}} \\
  & tr         & lv         & lt  & Avg.          \\ \midrule
$\text{Base}_{char}$            &  61.79      & 61.18    & 39.39  & 54.12    \\ 
$\text{Base}_{phone}$           &  54.50      & 45.50  & 43.64 & 47.88    \\ 
$\text{Base}_{attr}$ (ours)     & \textbf{48.33} & \textbf{40.10}  & \textbf{30.30} & \textbf{39.58} \\ \midrule \midrule
$\text{DAT}_{char}$             &  61.23      & 57.33     & 45.45     & 54.67    \\ 
$\text{DAT}_{phone}$            & 46.67      & 39.59     & 49.70     & 46.33    \\ 
$\text{DAT}_{attr}$ (ours)      & \textbf{42.96} & \textbf{36.25}  & \textbf{26.36} & \textbf{37.10} \\ \bottomrule 
\end{tabular}
\end{adjustbox}
\vspace{-.4cm}
\end{table}

\vspace{-.05cm}
\subsubsection{Zero-shot Transfer: OOV amd Unseen Languages}
\vspace{-.05cm}
Table~\ref{tab:idoov_results} and~\ref{tab:ul_results} show the WER for the out-of-domain scenarios, namely in-domain out-of-vocabulary (\textbf{ID OOV}) and unseen languages (\textbf{UL}). For OOV keywords, character-based system degrades significantly, due to the fact that some of the character tokens are not included during training. However, phoneme-based systems, namely $\text{Base}_{phone}$ and $\text{DAT}_{phone}$, perform slightly better than $\text{Base}_{attr}$ and $\text{DAT}_{attr}$ but the improvements are not meaningful. We argue that since all the tokens are still covered in the phoneme set, both phoneme and attribute systems have similar results. 

For unseen languages, our proposed framework $\text{Base}_{attr}$ consistency outperforms $\text{Base}_{char}$, $\text{Base}_{phone}$ in all three languages, with relative WER reduction of 26.87\% and 17.34\% respectively. Comparing DAT and Baseline, we can see that DAT improves the robustness across languages of both phonemes and attributes systems, and slightly degrade the character system on average. However, the attribute-based system clearly benefits more from the language-invariant features generated by the DAT framework as we can see all three unseen languages yield meaningful improvements, while character- and phoneme-based systems degrade on Lithuanian. Furthermore, a larger reduction on WER (from 39.58\% to 37.10\%) can be achieved. Comparing DAT framework on the three tokens, a similar pattern as Baseline can be observed where $\text{DAT}_{attr}$ consistently outperforms both $\text{DAT}_{char}$ and $\text{DAT}_{phone}$, with relative WER reduction of 32.14\% and 19.92\%, respectively.

\vspace{-.1cm}
\section{Conclusion}
\label{sec:conclusion}
\vspace{-.1cm}
A novel system to build a language-universal multilingual SKR system leveraging pre-trained self-supervised models and speech attribute modeling has been presented.  Our findings strongly confirm that speech attributes are a viable solution toward building multilingual speech recognition models. In fact, the proposed solution has the capability of generalizing toward out-of-domain scenarios, such as out-of-vocabulary keywords and unseen languages. We also observed significant word error rate reduction under zero-shot transfer to three low-resource languages, namely, Turkish, Latvian, and Lithuanian. Meanwhile, it is imperative to emphasize that our system has exceptional flexibility and seamless compatibility with various advanced training techniques and large pre-trained models. This adaptability highlights its potential for diverse applications and sets a foundation for future integration toward a unified framework for multilingual ASR.

\footnotesize

\bibliographystyle{IEEEtran}
\bibliography{mybib}

\end{document}